\documentclass[aps,prl,twocolumn,superscriptaddress]{revtex4-2}

\usepackage[utf8]{inputenc}
\usepackage{siunitx}  
\usepackage{dirtytalk}
\usepackage{csquotes}
\usepackage[version=4]{mhchem}
\usepackage{verbatim}
\usepackage{upgreek}  
\usepackage{changepage}
\usepackage{braket}
\usepackage[hidelinks]{hyperref}
\usepackage{graphicx}
\usepackage{animate}

\newcommand{\Ub}{U_{Bias}}

\begin{document}

\title{Temperature dependence of surface superconductivity in \ce{t-PtBi2}}

\author{Julia Besproswanny}
\email[]{besproswanny@uni-wuppertal.de}

\author{Sebastian Schimmel}

\affiliation{Fakultät für Mathematik und Naturwissenschaften, Bergische Universität Wuppertal, Gaußstraße 20, 42119 Wuppertal, Germany}
\affiliation{Leibniz-Institute for Solid State and Materials Research (IFW-Dresden), Helmholtz Straße 20, 01069 Dresden, Germany}

\author{Yanina Fasano}
\affiliation{Instituto de Nanociencia y Nanotecnología and Instituto Balseiro, CNEA – CONICET and Universidad Nacional de Cuyo, Centro Atómico Bariloche, Avenida Bustillo 9500, 8400 Bariloche, Argentina}
\affiliation{Leibniz-Institute for Solid State and Materials Research (IFW-Dresden), Helmholtz Straße 20, 01069 Dresden, Germany}

\author{Grigory Shipunov}
\affiliation{Leibniz-Institute for Solid State and Materials Research (IFW-Dresden), Helmholtz Straße 20, 01069 Dresden, Germany}
\affiliation{Current address: Institute of Physics, University of Amsterdam, 1098 XH Amsterdam, The Netherlands}

\author{Saicharan Aswartham}
\affiliation{Leibniz-Institute for Solid State and Materials Research (IFW-Dresden), Helmholtz Straße 20, 01069 Dresden, Germany}
\affiliation{Current address: Department of Physics and Astronomy, Ruhr Universität Bochum, Universitätsstraße 150, 44801 Bochum, Germany}

\author{Danny Baumann}
\affiliation{Leibniz-Institute for Solid State and Materials Research (IFW-Dresden), Helmholtz Straße 20, 01069 Dresden, Germany}

\author{Bernd Büchner}
\affiliation{Leibniz-Institute for Solid State and Materials Research (IFW-Dresden), Helmholtz Straße 20, 01069 Dresden, Germany}
\affiliation{Institute of Solid State and Materials Physics and Würzburg-Dresden Cluster of Excellence ct.qmat, Technische Universität Dresden, 01062 Dresden, Germany}

\author{Christian Hess}
\email[]{c.hess@uni-wuppertal.de}
\affiliation{Fakultät für Mathematik und Naturwissenschaften, Bergische Universität Wuppertal, Gaußstraße 20, 42119 Wuppertal, Germany}

\date{\today}

\begin{abstract}

The Weyl semimetal trigonal \ce{PtBi2} has recently been identified as a promising candidate material for intrinsic topological surface superconductivity emerging from the Fermi arc states of the material with a sizeable superconducting gap. We report the temperature evolution of the superconducting excitation spectrum using scanning tunneling spectroscopy in the range of $\SIrange{8}{45}{\kelvin}$. 
A large low-temperature gap in the order of $\Delta \approx \SI{9}{\milli\electronvolt}$ and a closing of the gap around $T_c \approx \SI{45}{\kelvin}$ is observed.
Thus, our results confirm the previously indicated high $T_c$-like superconductivity in \ce{t-PtBi2}.

\end{abstract}

\maketitle

The time reversal invariant Weyl semimetal t-\ce{PtBi2} \cite{Kuibarov2024, hoffmann2024, Veyrat2023}, has recently attracted considerable attention because of unusual surface superconductivity \cite{Kuibarov2024, Schimmel2024}. 
Bulk resistivity on macroscopic crystals as well as on nano-flakes reveal transitions towards zero resistance only at very low temperatures of $\SIrange{0,6}{1,1}{\kelvin}$ \cite{shipunov2020, zabala2024}. In contrast, the surface sensitive techniques scanning tunneling spectroscopy (STS) and angular-resolved photoemission spectroscopy (ARPES) report sizable superconducting gaps even at much higher $T\approx\SI{4}{\kelvin}$ \cite{Schimmel2024, Kuibarov2024}.
Since the surface electronic structure of Weyl semimetals is expected to carry characteristics of the bulk Weyl nodes projected onto the surface in the form of so-called Fermi arcs, one might conjecture that this surface superconductivity emerges from these topological Fermi arc states. Notably, ARPES reports sharp signatures of the superconducting gap  of the order $\Delta\approx\SIrange{1}{2}{\milli\electronvolt} $ to be located in reciprocal space precisely at the position of the Fermi arcs, and a critical temperature of $\SIrange{8}{1}{\kelvin}$ \cite{Kuibarov2024}. On a more local scale, STS reports even larger gap sizes up to about $\SI{20}{\milli\electronvolt}$ at $\SI{5}{\kelvin}$, implying even higher critical temperatures \cite{Schimmel2024}.
Here, we report the temperature evolution of such a gap by means of STS. More specifically, we explore a surface area of t-\ce{PtBi2} that exhibits a comparatively large gap of approximately $\SI{9}{\milli\electronvolt}$ in the range from $\SIrange{8}{45}{\kelvin}$.  With a $T_c \geq \SI{45}{\kelvin}$, our data corroborate the expectation of high $T_c$ surface superconductivity of \ce{t-PtBi2} deduced from the energy scale of the gap measured previously at 5K.

\begin{figure}[t]
	\includegraphics[width=0.38\textwidth]{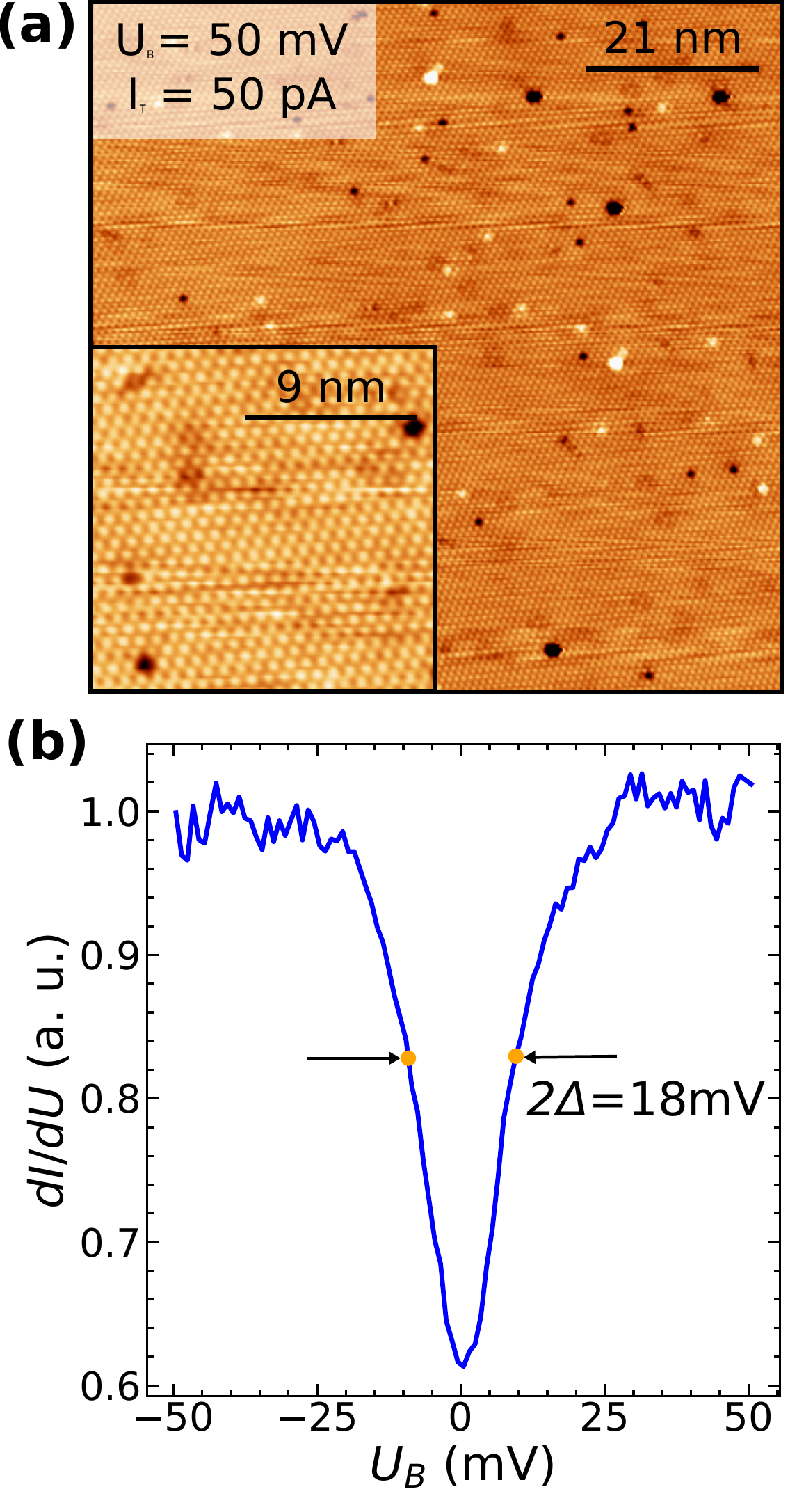}
	\caption{
		(a): Representative $76\times\SI{76}{\nano\meter^2}$ topography of the studied \ce{PtBi2} surface measured at $T = \SI{18}{\kelvin}$ with set point parameters $\Ub = \SI{50}{\milli\volt}$ and $I_T = \SI{50}{\pico\ampere}$. Inset: digital zoom of an $18.5\times\SI{18.5}{\nano\meter^2}$ area. The image shows a clear Type A termination (see classification in \cite{Schimmel2024}). (b): $dI/dU$ spectrum of \ce{PtBi2} at $T = \SI{8}{\kelvin}$. It was obtained with set point parameters $\Ub = \SI{50}{\milli\volt}$ and $I_{T} = \SI{500}{\pico\ampere}$ as an average of $16$ point spectra along a $4 \times 4$ grid on an $80\times\SI{80}{\nano\meter^2}$ area. The orange points and black arrows indicate the $FWHM$ value. $FWHM/2$ is used as an estimate for the gap size $\Delta_{FWHM/2} = \SI{9\pm0.5}{\milli\electronvolt}$
	}
	\label{fig:topo18K_8K_spec}
\end{figure}

The experiment was conducted on a t-\ce{PtBi2} single crystal grown via the self-flux method \cite{shipunov2020}.
The measurements have been performed in a home-built liquid Helium-cooled scanning tunneling microscope operated by a Nanonis SPM control system. We regulated the sample temperature in a range of $T = \SIrange{8}{45}{\kelvin}$ by applying current to a heater located in the STM head. The setpoint temperature was stabilized by a PID controller. \\
A mechanically sharpened \ce{PtIr} wire was used as the scanning tip. To obtain an atomically clean sample surface, the t-\ce{PtBi2} crystal was cleaved in cryogenic ultra high vacuum at a temperature of $T \approx \SI{8}{\kelvin}$. Topographic data was acquired by scanning the sample surface in constant current mode and subsequently processing the data using the WSxM software \cite{wsxm}. Differential conductance ($dI/dU$) spectra were obtained through the commonly used lock-in technique with $f_{mod} = \SI{1.1}{\kilo\hertz}$ and $U_{mod} = \SI{2}{\milli\electronvolt}$. The integration time per data point was set between $\SIrange{5}{10}{\milli\second}$ and each presented spectrum is an average of $16$ point spectra in a $4\times 4$ grid with an average over $10 - 40$ sweeps per point spectrum. The setpoint conditions for each spectrum were set to bias voltage $\Ub = \SI{50}{\milli\volt}$ and tunneling current $I_T = \SI{500}{\pico\ampere}$.

A representative topography for the sample under investigation is shown in \autoref{fig:topo18K_8K_spec} (a). The atomic corrugation features a triangular lattice, which, based on a comparison to topographies presented in previous STM studies \cite{Nie2020, Schimmel2024, hoffmann2024}, can be assigned to the decorated honeycomb (Type A) surface. 
Note that, as an inversion symmetry breaking van der Waals material, t-\ce{PtBi2} exhibits two distinct \ce{Bi}-terminated surfaces. On one side of the crystal the \ce{Bi} surface atoms are in a coplanar arrangement, denoted as Kagome-like (or Type B) surface. The opposite surface, at which one \ce{Bi} atom of each unit cell is slightly lifted out of the plane, is referred to as decorated honeycomb (or Type A) \cite{Nie2020}. 
The ideal decorated honeycomb atomic corrugation is only interrupted by localized point defects of low density $\approx \SI{1}{\percent}$ per unit cell, displaying the high quality of the single crystal.

The differential tunneling conductance, as measured by STS, is proportional to the sample electronic density of states convoluted with the width of the Fermi edge, thus providing direct access to the superconductor's excitation spectrum characterized by the superconducting energy gap. 
A normalized differential conductance ($dI/dU$) spectrum representative for the electronic structure of the investigated area at $T = \SI{8}{\kelvin}$ is presented in \autoref{fig:topo18K_8K_spec} (b). The spectrum features a clear gap with a zero bias conductance (ZBC) reduction by $\SI{38}{\percent}$ centered at the Fermi level, in agreement with our previous findings on different superconducting t-PtBi2 surfaces \cite{Schimmel2024}. The nonzero ZBC value can be attributed to the surface nature of the superconductivity, with the bulk remaining in normal state \cite{Schimmel2024, Kuibarov2024}. Like occasionally observed before \cite{Schimmel2024}, sharp coherence peaks are absent.

Due to the lack of coherence peaks we refrain from the exact determination of the gap size $\Delta$ based on a fit with a BCS density of states \cite{Dynes}. Instead, we estimate $\Delta_{est}$ at $T = \SI{8}{\kelvin}$ by using the half width at half minimum ($FWHM/2$) value \cite{footnote}. 
With this approach we determine  $\Delta_{est} = FWHM/2 \approx \SI{9\pm0,5}{\milli\electronvolt}$. Assuming in a first approximation that t-\ce{PtBi2} is a conventional weakly coupled BCS superconductor, the BCS ratio $\Delta_0 = 1.76 k_B T_c$ suggests a $T_c \approx \SI{60}{\kelvin}$. Motivated by this high value of $T_c$, we therefore study the temperature evolution of the gap up to the highest possible temperature while maintaining consistent tunneling conditions.

\begin{figure}
	\includegraphics[width=0.45\textwidth]{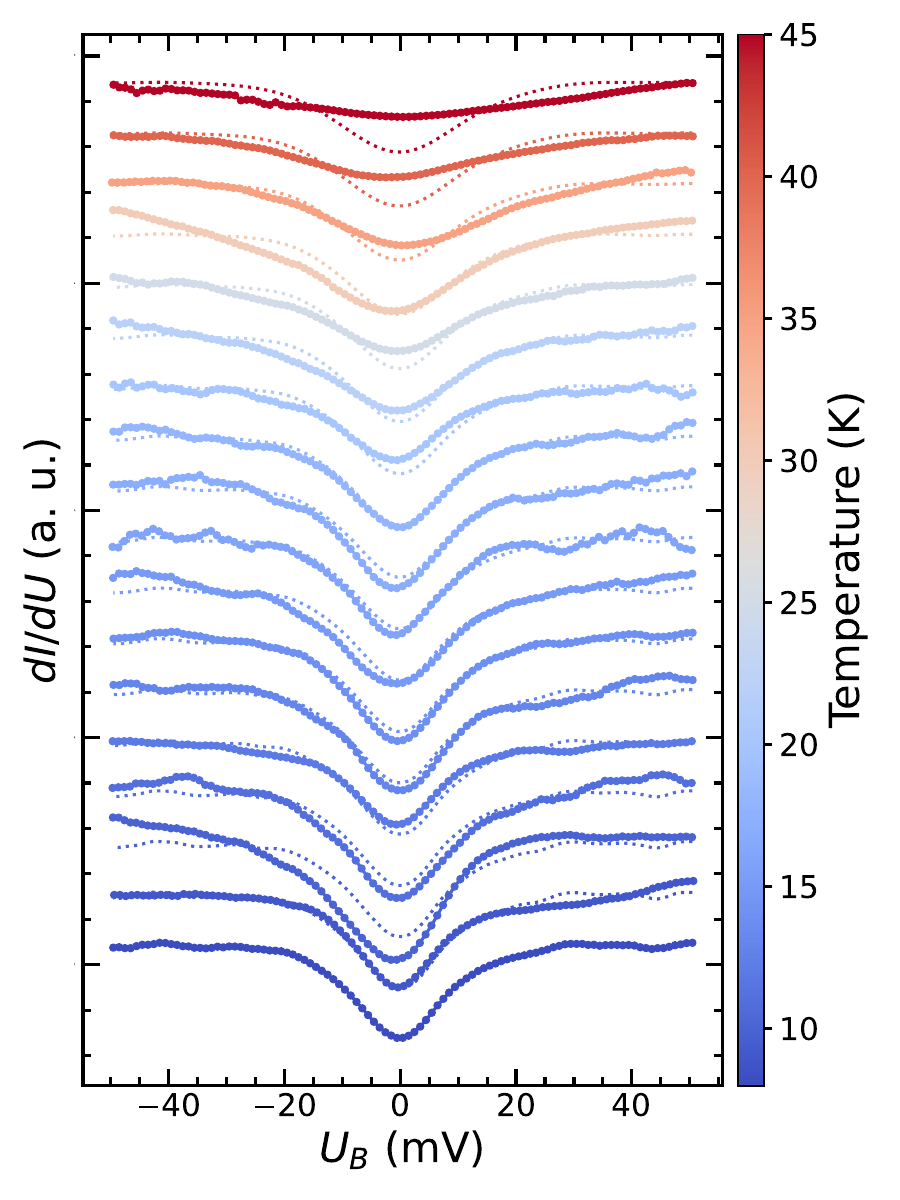}
	\caption{
		Evolution of the $dI/dU$ spectra measured via lock-in technique at temperatures from $\SI{8}{\kelvin}$ to $\SI{45}{\kelvin}$. The spectra have been smoothed using a moving average over 6 points and are offset for clarity. The results are obtained after subtracting a linear fit in the entire energy range from each spectrum \cite{supplement}. The dotted lines represent the convolution of the $\SI{8}{\kelvin}$ spectrum with the Fermi function derivative for the corresponding temperatures.
	}
	\label{fig:waterfall}
\end{figure}

\begin{figure}
	\includegraphics[width=0.39\textwidth]{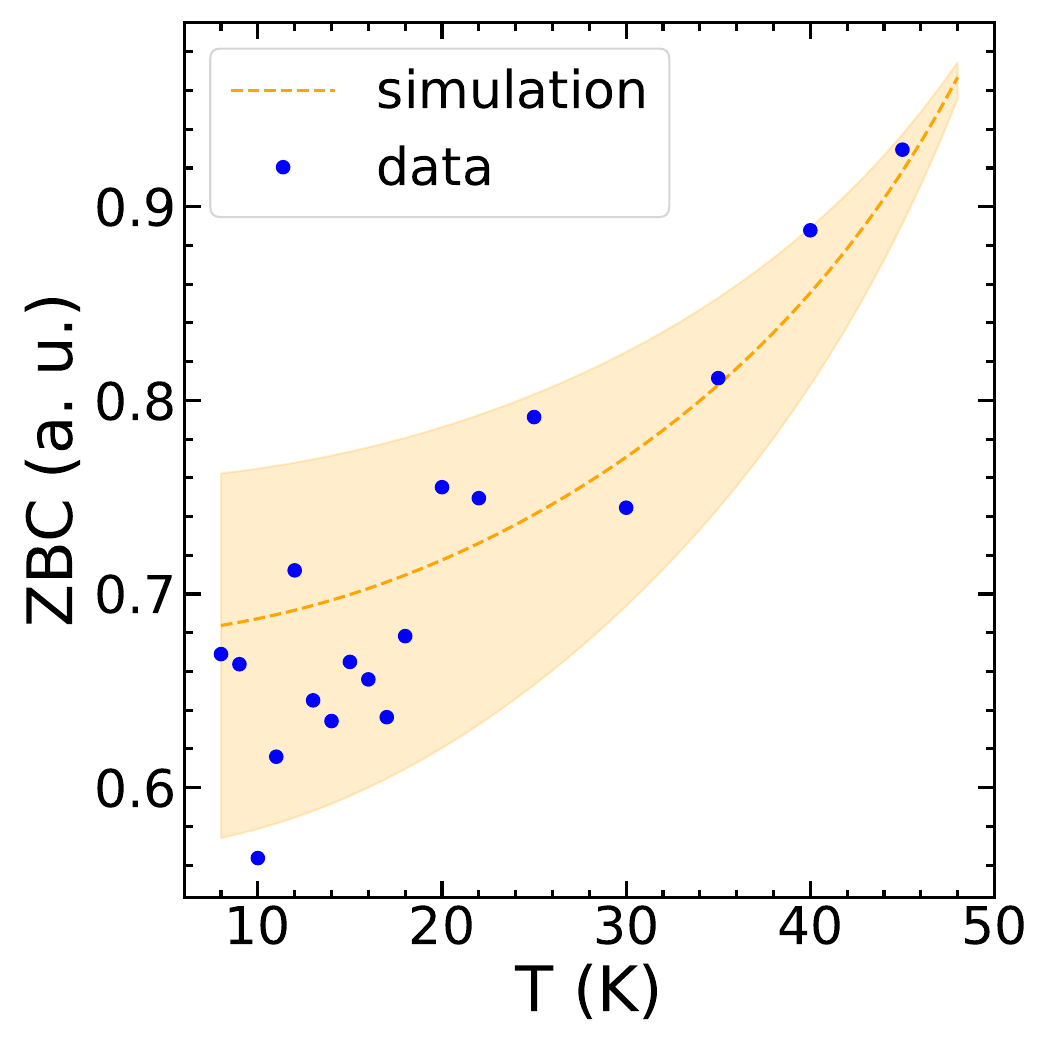}
	\caption{Evolution of the ZBC with temperature. The blue circles show the ZBC obtained from the data set presented in \autoref{fig:waterfall}. The simulated curve (dashed orange line) was obtained from simulations assuming an s-wave ordered gap with a size of $\Delta = \SI{9}{\milli\electronvolt}$, a Dynes parameter $\Gamma = \SI{8\pm2}{\milli\electronvolt}$ and a BCS-ratio $R_{BCS} = 4.2$ (for more details, see Supplemental Material \cite{supplement}). The orange background around the simulated curve shows the margins for $\Gamma = 6$ and $\Gamma = 10$. The curves show a qualitatively similar behavior on a comparable numerical scale.}
	\label{fig:ZBC}
\end{figure}

\autoref{fig:waterfall} shows the central result of this study, i.e., the STS average spectra obtained on warming up to $\SI{45}{\kelvin}$ (solid lines).
As can be clearly seen, the energy gap at the Fermi level shows a clear evolution with temperature: Upon increasing the temperature, the reduction of the ZBC lifts at an increasing rate, until the gap is almost completely suppressed at $T = \SI{45}{\kelvin}$ \cite{supplement}. 

We emphasize that the observed temperature evolution of the spectra can be ruled out to simply result from trivial thermal broadening. 
To demonstrate this fact, the thermally broadened $\SI{8}{\kelvin}$ spectrum is superimposed on the measured spectra for each corresponding temperature (dashed lines in \autoref{fig:waterfall}). The intrinsic nature of the gap closing is particularly well visible when comparing the measured spectra (solid lines in \autoref{fig:waterfall}) at $\SI{40}{K}$ and $\SI{45}{K}$, which feature only a very small reduction of the ZBC, with the broadened $\SI{8}{\kelvin}$ spectra, where the dip is persisting. Hence, the closing of the gap can be attributed to a real change of the electronic structure -- the transition between the superconducting ground state and the normal state at elevated temperatures. 
The gap closing is further validated by the temperature evolution of the ZBC as presented in \autoref{fig:ZBC}. As mentioned before, the gap magnitude is not fitted with a BCS fit due to the lack of clear coherence peaks. Therefore, a temperature evolution of $\Delta(T)$ cannot reliably be extracted. Nevertheless, the temperature dependence of the ZBC values follows the expectation for a superconductor with large life-time broadened spectra, as illustrated in \autoref{fig:ZBC} (for further details, see Supplemental Material \cite{supplement}).

Based on our observations, the critical temperature of this surface can be estimated to $T_c \approx \SI{50\pm5}{\kelvin}$, closely above our highest consistently measurement temperature of $T = \SI{45}{K}$. At $T = \SI{50}{\kelvin}$, there is no depletion of the ZBC, as seen in Supplemental Material \cite{supplement} Figure 3. Such a high transition temperature is to date only known for unconventional superconductors like cuprates and Fe-based superconductors \cite{Buckel2012, Stewart2011}. Considering the estimated transition temperature and the determined $\Delta_{est} = \SI{9\pm0,5}{\milli\electronvolt}$, the BCS ratio results in $R_{BCS} = \frac{2\Delta}{k_BT_c} = 4.2\pm0.2$. This value is only slightly higher than $R_{BCS} = 3.54$ of weak-coupling superconductors and is well in line with conventional superconductors of which many exhibit even higher ratios. E.g., one finds \ce{Hg} with $R_{BCS} = 4.6$ and \ce{MgB2} with up to $R_{BCS} = 4.5$ \cite{Buckel2012}, while unconventional cuprate superconductors yield values significantly higher (see \cite{fischer2007scanning} for an overview). 
Taking into account the $T_c$, which is higher than in any known conventional superconductor including \ce{MgB2}, and the BCS ratio, surface superconductivity in t-\ce{PtBi2} might be classified as a weak- to moderate-coupling high-$T_c$, unlike cuprate superconductors, which are characterized by strong coupling \cite{fischer2007scanning}. A more precise determination of $\Delta$ is required to further validate this conjecture. Note further, that it is possible that the $T_c$ is influenced by the two-dimensionality of the surface superconductivity: A reduction \cite{Ugeda2016} and an increase \cite{NavarroMoratalla} of $T_c$ have both been observed in thin layers of bulk superconductors. \\
Notably, no coherence peaks can be identified in the $dI/dU$ spectra. This unexpected behavior has as occasionally been observed also in Fe- or Cu-based superconductors \cite{kawashima2015, Kashiwaya1998, tanaka1994}. The origin of this phenomenon in \ce{t-PtBi2} is unclear. One might speculate, that superconducting order parameter fluctuations due to incoherent pair formation \cite{Raychaudhuri2022} or proximity to a quantum critical point \cite{Roy2018,Ienaga2020}, both often appearing in low dimensional or disordered superconductors \cite{Raychaudhuri2022}, can lead to the formation of a gap with partially or completely suppressed coherence peaks \cite{Norman2005}. 
These phenomena have also been hypothesized to be the source of pseudo gap phases above $T_c$ in, e.g., cuprates \cite{Damascelli,fischer2007scanning,Norman2005} and twisted bilayer graphene \cite{Oh2021}.
This however does not explain the discrepancy with particularly sharp coherence peaks from ARPES observations \cite{Kuibarov2024} and in some of our previously reported STS data \cite{Schimmel2024}. This might be connected to a key difference between STS and ARPES: While ARPES is a momentum space-resolved measurement technique, STS has no momentum space resolution. Instead STS averages over many different $k$ states. The sensitivity of the tunneling junction for different k-space momenta, described by the tunneling matrix element, can vary for different tip-to-surface systems. Since the tunneling matrix element in an STS experiment cannot be controlled deliberately, a corresponding influence on the measured spectra cannot be excluded \cite{fischer2007scanning}. ARPES results suggest that superconductivity in this compound occurs on the very localized crystal momenta of the Fermi arcs. Since the Fermi arcs are located considerably away from the $\Gamma$ point, one might speculate a profound influence of the tunneling matrix element. The potential existence of nodes \cite{Schimmel2024} in this already very localized gap further complicates this influence.

In conclusion, we present a temperature dependent study of the electronic structure of topological Weyl semimetal and surface superconductor \ce{t-PtBi2}. The density of states spectra clearly show an energy gap with an estimated size of $\Delta_{est} = \SI{9\pm0,5}{\milli\electronvolt}$. A gradually increased closing of the gap can be observed up to $\SI{45}{\kelvin}$, thus the critical temperature $T_c$ can be estimated unconventionally high at around $T_c \approx \SI{50\pm5}{\kelvin}$. We find a BCS ratio $R_{BCS} = \frac{2\Delta}{k_BT_c} = 4.2\pm0.2$, close to that of a weakly-coupled superconductor $R_{BCS} = 3.54$. The results provide further evidence for the exotic surface superconductivity in t-\ce{PtBi2}, which was previously proposed to be high $T_c$ based on the magnitude of the gap and is hereby confirmed by temperature dependent STS measurements.
With the combination of high $T_c$ superconductivity and a topologically non-trivial electronic structure, t-\ce{PtBi2} qualifies as a potential easy access intrinsic topological superconductor. Such rare systems are considered a playground for novel quantum phenomena which may host zero bias Majorana modes. These are believed to constitute a key ingredient towards fault-tolerant quantum computing \cite{Kitaev_2001}.

\section{Acknowledgments}
This work received support from the Deutsche Forschungsgemeinschaft (grants 500507880, 566479091 and SFB 1143) as well as the Dresden–Würzburg Cluster of Excellence (EXC 2147). Furthermore, this project received funding from the European Research Council (grant 647276 – MARS – ERC-2014-CoG). Y.F. acknowledges support from the Alexander von Humboldt Stiftung and from the Dresden Technische Universität Senior Fellowship Program.

\bibliography{refs}

\newpage

\widetext

\setcounter{equation}{0}
\setcounter{figure}{0}
\setcounter{table}{0}
\makeatletter
\renewcommand{\theequation}{S\arabic{equation}}
\renewcommand{\thefigure}{S\arabic{figure}}
\renewcommand{\bibnumfmt}[1]{[S#1]}
\renewcommand{\citenumfont}[1]{S#1}

\section{Supplementary information}

\section{Gap simulations}

\begin{figure*}[b]
	\includegraphics[width=0.9\textwidth]{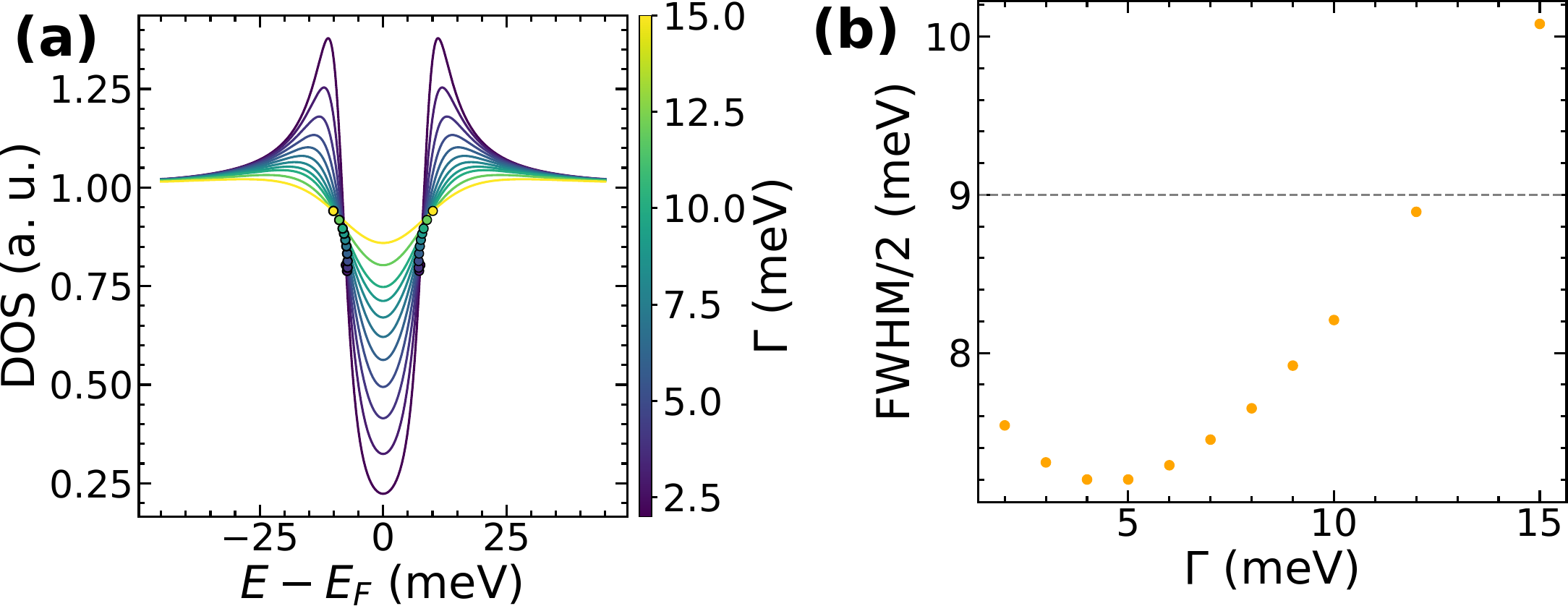}
	\caption{Simulation results of the FWHM value on an s-wave gap using a BCS density of states with a Dynes parameter accounting for a shortening of quasiparticle lifetime \eqref{eq:dynes}. (a): DOS for a gap of size $\Delta = \SI{9}{\milli\electronvolt}$ for different lifetime parameters $\Gamma$. Additionally, each curve is convolved with the derivative of the Fermi function at $\SI{8}{\kelvin}$. The circles indicate the FWHM position for each spectrum. (b): The half FWHM value obtained from the curves in (a) plotted against $\Gamma$. The horizontal dashed gray line indicates the simulated gap size $\Delta = \SI{9}{\milli\electronvolt}$}
	\label{fig:FWHM_sim}
\end{figure*}

\autoref{fig:FWHM_sim} shows a simulation of the energy gap of a superconductor with $\Delta(0)=\SI{9}{\milli\electronvolt}$ to illustrate the evolution of the full width at half minimum (FWHM) depending on the Dynes lifetime parameter $\Gamma$. The Dynes equation 

\begin{align}\label{eq:dynes}
	DOS(\varepsilon; \Gamma) = Re\left\{\frac{\varepsilon-i\Gamma}{(\varepsilon-i\Gamma)^2-\Delta(T)^2}\right\}, \qquad \varepsilon = E - E_F
\end{align}

was used to calculate the curves for different values of the lifetime parameter $\Gamma=\SI{2}{\milli\electronvolt}$ to $\SI{15}{\milli\electronvolt}$ (\autoref{fig:FWHM_sim} (a)). Additionally, each spectrum is convolved with the derivative of the Fermi function at $\SI{8}{\kelvin}$ in order to take the thermal broadening into account.
With increasing $\Gamma$, the coherence peaks are increasingly suppressed. This is consistent with the missing of coherence peaks in the data. The circles in \autoref{fig:FWHM_sim} (a) indicate the calculated FWHM position for each spectrum. In \autoref{fig:FWHM_sim} (b) the half FWHM value is plotted against $\Gamma$ and the true $\Delta$ is indicated by the horizontal dashed gray line. It can be clearly seen that the $FWHM \leq \Delta$ up to a large $\Gamma = \SI{12}{\milli\electronvolt}$. Thus, for reasonable values of $\Gamma$ the FWHM value provides a lower bound of the true gap size $\Delta$.

\begin{figure*}
	\includegraphics[width=0.9\textwidth]{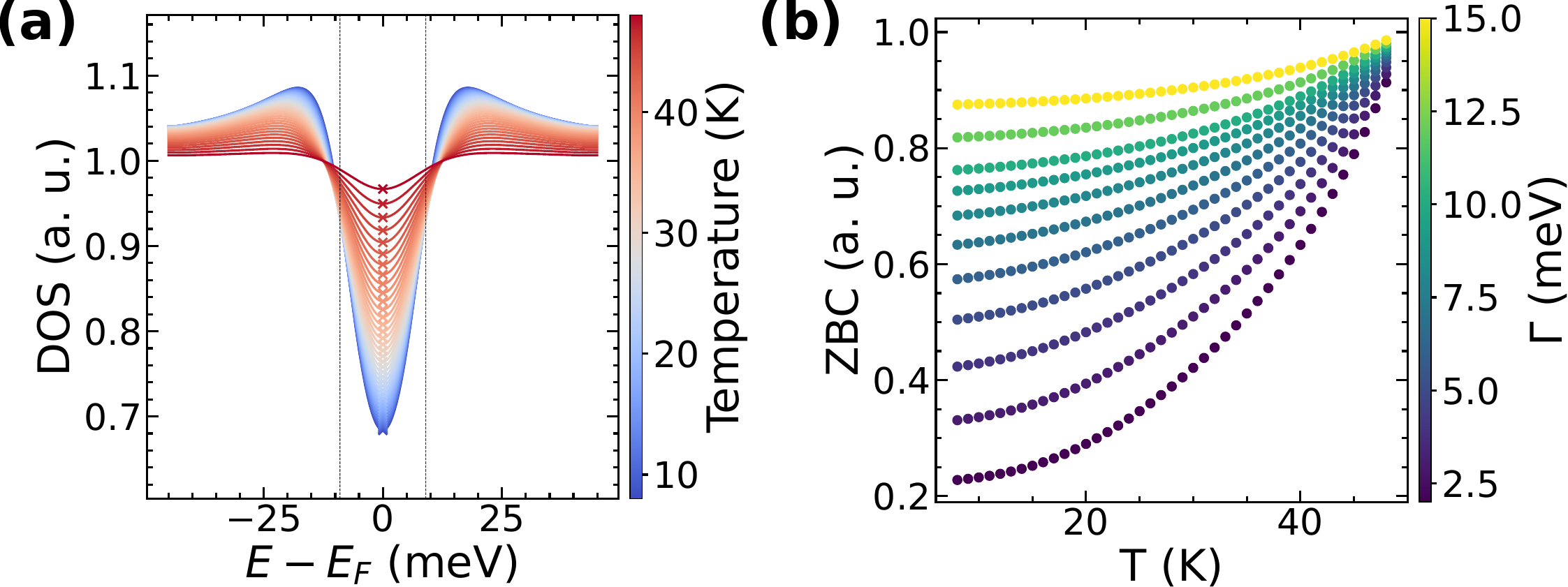}
	\caption{Simulation results of the ZBC on an s-wave gap using a BCS density of states with a Dynes parameter accounting for a shortening of quasiparticle lifetime \eqref{eq:dynes}. (a): DOS for a gap of size $\Delta = \SI{9}{\milli\electronvolt}$ and lifetime parameter $\Gamma = \SI{8}{\milli\electronvolt}$. The temperature evolution of the gap is simulated via the convolution with the derivative of the Fermi function at the corresponding temperatures. The crosses indicate the ZBC position for each spectrum. (b): For each $\Gamma$ the ZBC values obtained from the curves in (a) are plotted against $T$. The curves are qualitatively similar.}
	\label{fig:ZBC_sim}
\end{figure*}

The temperature evolution of the zero bias conductance (ZBC) for different $\Gamma$ is presented in \autoref{fig:ZBC_sim}. In (a), the DOS for $\Gamma = \SI{8}{\milli\electronvolt}$ is plotted as a representative example for different temperatures with crosses marking the ZBC. \autoref{fig:ZBC_sim} (b) shows the temperature dependence of the ZBC obtained in an equivalent manner for various $\Gamma = \SIrange{2}{15}{\milli\electronvolt}$. The curves show a qualitatively similar progression on different numerical scales. 

\section{Spectroscopy data}

\begin{figure*}
	\includegraphics[width=0.9\textwidth]{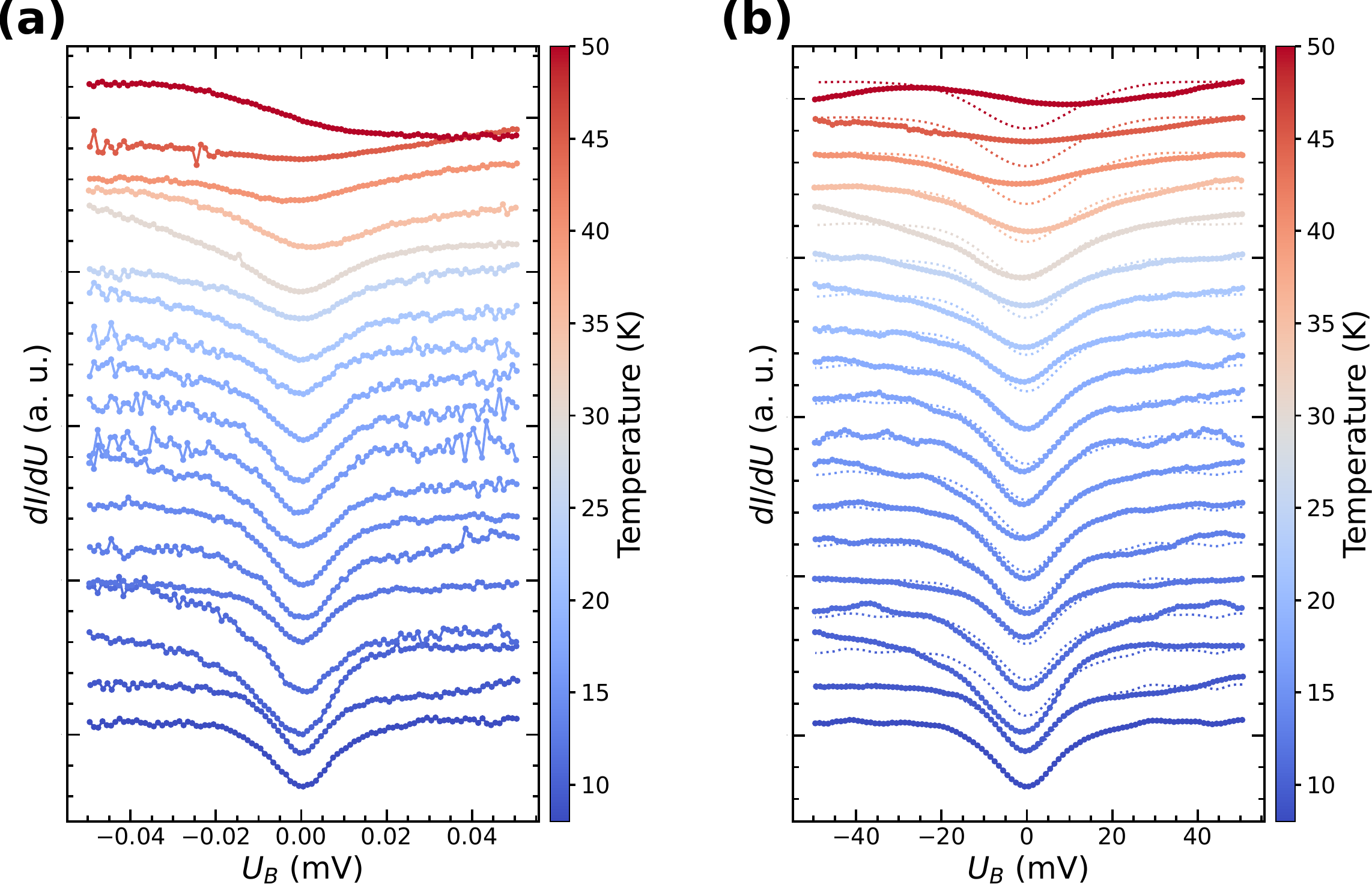}
	\caption{Comparison of the raw and processed $dI/dU$-spectra measured via the lock-in method in a temperature range $\SIrange{8}{50}{\kelvin}$. (a): $dI/dU$-spectra obtained by averaging over 16 point spectra in a $4\times4$ grid. The spectra are normalized and shifted along the $y$-axis for clarity. (b): the same spectra after a linear fit is performed on each spectrum and the resulting slope is subtracted. Smoothing is applied using a moving average over 6 points. The dotted lines superimposed on the measured curves represent the convolution of the $\SI{8}{\kelvin}$ spectrum with the derivative of the Fermi function for the corresponding temperature.}
	\label{fig:data_raw}
\end{figure*}

The following \autoref{fig:data_raw} serves as a comparison between the measured $dI/dU$ spectra in a temperature range $T = \SIrange{8}{50}{\kelvin}$ before and after the data analysis steps described in the main text. \autoref{fig:data_raw} (a) presents the data obtained by the lock-in method, averaged over 16 point spectra from a $4 \times 4$ grid and normalized. Each of these curves is then fitted with a linear fit to extract and subtract the slope, in order to exclude a changing linear background. Potential sources of such a slope are a contribution from the normal state to the DOS and a drift of the STM tip along the $z$ axis, e. g. due to heating above the base temperature of the STM. The temperature relaxation of the sample and tip can differ, increasing the thermal drift during the measurement. Additionally, the spectra are smoothed by a moving average over 6 points. The result after both steps is shown in \autoref{fig:data_raw} (b). As in the main text, the dotted lines are superimposed on the measured spectra to illustrate the difference between the simple thermal broadening of the gap and its closing.	

Notably, this figure features a spectrum obtained at $T = \SI{50}{\kelvin}$ which was not presented in the main text. In \autoref{fig:data_raw} (a) one can see a much more significant change in slope and shape, which we attribute to a change of the tunneling junction. Thus, this spectrum is not considered to be consistent with the previous spectra. Despite this, it is still noted that it does not include a gap, but rather a step-like anomaly near the Fermi level. We conclude, that at $T = \SI{50}{\kelvin}$ the gap is fully closed.

\end{document}